 \documentclass[final,1p,times,twocolumn]{elsarticle}

\usepackage{amssymb}

\journal{Chemical Physics Letters}

\begin{document}


\begin{frontmatter}

\title{Adsorption of the astatine species on gold surface: a relativistic density functional theory study}

\author[ad1,ad2]{Yuriy Demidov}
\ead{iurii.demidov@gmail.com}

\author[ad1,ad3]{Andr\'ei Zaitsevskii}

\address[ad1]{National Research Centre ``Kurchatov Institute'' B.P. Konstantinov Petersburg Nuclear Physics Institute,\\ Orlova Roscha, 188300 Gatchina, Russia}
\address[ad2]{St. Petersburg Electrotechnical University ``LETI'', Physics Department, Prof. Popov 5, 197376 St. Petersburg, Russia}
\address[ad3]{M.V.~Lomonosov Moscow State University,Chemistry Department, Vorob'evy gory 1, 119991 Moscow, Russia}

\begin{abstract}
We report on first-principle based studies of the adsorption interaction of astatine species on a gold surface. 
These studies are aimed primarily at the support and interpretation of gas chromatographic experiments with Superheavy Elements, tennessine (Ts, $Z=117$) 
as the heavier homologue of At and possibly nihonium (Nh, $Z=113$) as its pseudo-homologue.
The adsorption energies of elemental astatine or the corresponding monohydroxide on a stable gold (111) surface are estimated using gold clusters with up to 69 atoms 
in order to simulate the adsorption site.
To simulate the electronic structure of $\rm At-Au_n$ and $\rm AtOH-Au_n$ complexes, we combine accurate shape-consistent relativistic pseudopotentials 
and non-collinear two-component relativistic density functional theory. 
The predicted adsorption energies for At and AtOH on gold are $\rm 130 \pm 10$~kJ/mol and $\rm 90 \pm 10$~kJ/mol, respectively.\end{abstract}


\end{frontmatter}

\section*{Introduction}
The discovery of relatively long-lived isotopes of Superheavy Elements (SHEs) in $\rm ^{48}Ca$-induced nuclear fusion reactions~\cite{ogan_rew} shows that the shore of the island of particularly shell-stabilized nuclei has indeed been reached~\cite{island,og_island}. 
While the recent addition of four new elements to the Periodic Table~\cite{IUPAC} is a fantastic discovery in itself, the long half-lives of certain isotopes open up the possibility for chemical investigations. 
Hence, after the successful chemical characterization of elemental copernicium~\cite{exp_112aun} (Cn, $Z=112$) and flerovium~\cite{exp_114aun} (Fl, $Z=114$), nihonium (Nh, $Z=113$) 
increasingly catches the chemist's interest. First experimental results have been recently obtained at the Flerov Laboratory of Nuclear Reaction in Dubna, Russia~\cite{e113_exp, e113_exp_new}. 

On-line gas-phase thermochromatography is an unique method for chemically studying the usually very short-lived SHEs at a one-atom-at-a-time regime~\cite{zvara,exp_112aun}. 
These experiments are extremely sophisticated and, moreover, they are confronted with rather low statistics for determining the corresponding chemical properties of the SHE in question. 
Thus, reliable preliminary theoretical modeling is essential for the experimental success as well as for the correct and detailed interpretation of the available data.

From a theoretical standpoint, investigations on the properties of SHEs are especially challenging as they require a profound understanding of the electronic
structure in the presence of the strong fields imposed by the heavy nuclei~\cite{our_rew}. 
Relativistic effects may lead to dramatic dissimilarities in the chemical behavior of SHEs when comparing them to their lighter homologs. 
The calculated adsorption energy for single atoms of nihonium on a gold surface~\cite{rus113} ($\rm E^{Au}_{ads}(Nh) = 105\pm 10$~kJ/mol) differs substantially from the experimentally measured adsorption energy on gold of its nearest homolog, i.e., thallium~\cite{tl_aun}, ($\rm E^{Au}_{ads} (Tl) = 262 \pm 6$~kJ/mol). 
This raises the question of what one may learn from experiments with the lighter homologs of nihonium in terms of
understanding the SHE's chemistry. Hence, finding a chemical pseudo-homolog appears to be a promising alternative~\cite{e113oh_our}.

It has been shown experimentally that the desorption temperatures and energies of Cn~\cite{exp_112aun} and Fl~\cite{exp_114aun} atoms
from a gold surface are fairy close to each other and overall lower than those of their immediate homologs
Hg and Pb, observed on the same surface. This confirms theoretical predictions concerning the electronic structure of
Cn and Fl atoms: the strong relativistic stabilization of the s- and $\rm p_{1/2}$-shells in both Cn ($6d^{10}7s^{2}$) and Fl ($6d^{10}7s^{2}7p^2_{1/2}$) results in a closed-shell character of the ground states of these atoms~\cite{subperiodic}. 
Due to this unique feature of the $\rm 7^{th}$ row of the Periodic Table, the electronic structure of a Nh atom can be
interpreted as a Fl atom with a hole in its closed $\rm 7p_{1/2}$-subshell~\cite{e113oh_our}.
This observation seems to render astatine a closer chemical ``relative'' of Nh in comparison to the formal homologue Tl. Thus,
At might be a plausible chemical species for test experiments aiming at finding the optimum experimental conditions for
further explorations of the Nh chemistry.

The non-observation of any decays of Nh in the latest chemical experiment~\cite{e113_exp_new} demands for a modification of the experimental approach.
Therefore, future experimental efforts may focus primarily on the efficient formation and detection of NhOH molecules -- most likely the chemical species observed in the early investigations\cite{e113_exp}.
A study of the electronic structure of NhOH and AtOH molecules has been carried out earlier at the combined one-component coupled cluster CCSD(T) 
and two-component relativistic functional theory (2c-RDFT) level~\cite{e113oh_our}. 
The therein calculated enthalpies for the elimination of the hydroxyl-group at 0~K for both of these molecules are fairly similar (188~kJ/mol for NhOH and 174~kJ/mol~for AtOH)
and much smaller than the corresponding value for Tl (319 kJ/mol).

In this work, we make an attempt to calculate the adsorption energies of single atoms of At and of AtOH molecules on a gold (111) surface. The results can be directly compared to the available experimental data, derived for both of these chemical systems~\cite{at_gold}.
\begin{figure}[htb]
\begin{minipage}[h!]{0.24\linewidth}
\center{\includegraphics[width=1\linewidth]{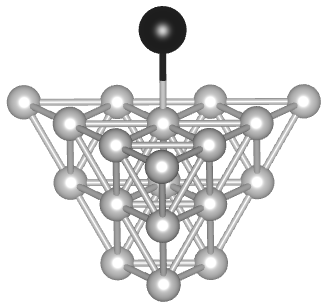}} \\
\end{minipage}
\hfill
\begin{minipage}[h!]{0.24\linewidth}
\center{\includegraphics[width=1\linewidth]{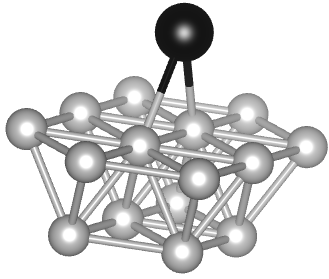}} \\
\end{minipage}
\hfill
\begin{minipage}[h!]{0.24\linewidth}
\center{\includegraphics[width=1\linewidth]{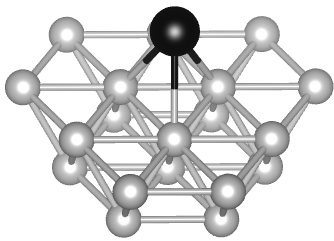}} \\
\end{minipage}
\hfill
\begin{minipage}[h!]{0.24\linewidth}
\center{\includegraphics[width=1\linewidth]{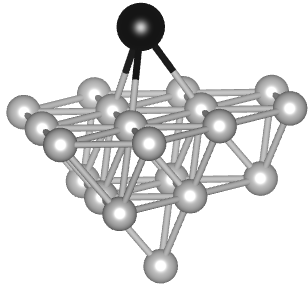}} \\
\end{minipage}
\begin{minipage}[h!]{0.24\linewidth}
\vfill
\center{on the top}
\end{minipage}
\hfill
\begin{minipage}[h!]{0.24\linewidth}
\center{bridge}
\end{minipage}
\hfill
\begin{minipage}[h!]{0.24\linewidth}
\center{hollow-2}
\end{minipage}
\hfill
\begin{minipage}[h!]{0.24\linewidth}
\center{hollow-3}
\end{minipage}
\caption{
The various types of considered $\rm At-Au_n$ complexes, corresponding to the different possible positions of the At atom (black sphere) bonded to one, two, or three gold atoms (gray spheres): 
``on the top'', ``bridge'', and ``hollow''. In the latter case, one has to discern the positions of the At atom with respect to the gold atom directly underneath in the second
(``hollow-2'') or third (``hollow-3'') layer from surface.
\label{fgr:pos}
}
\end{figure}

\section*{Method and computation details}
To simulate the electronic structure of the $\rm At/AtOH - Au_n$ adsorption complexes, we combine accurate shape-consistent relativistic pseudopotentials (RPPs) 
and non-collinear 2c-RDFT with two qualitatively different approximations for the exchange-correlational functional, i.e., the generalized-gradient B88P86~\cite{becke,perdew} 
and the partially non-empirical hybrid PBE0\cite{pbe0}.
The RPP model allows to replace the rather complicated problem of the relativistic interacting electron gas in the strong nuclear field by a much simpler problem  defined by the RPPs
treating the state of the electron gas in the external field as formally non-relativistic, all at the cost of a rather complex external field structure.
This replacement is accompanied by a sharp reduction of the number of variables (i.e., the wave functions or electron densities depend only on the coordinates
of the valence electrons) and by the elimination of the difficulty to approximate oscillations of the wave function in the vicinity of the atomic nuclei.
RPPs were used for the representation of 60 core electrons of both, Au~\cite{gibrid} as well as At~\cite{e113oh_our}.
Meanwhile, the outer-core electrons, i.e., 19 for Au and 25 for At, were explicitly described within the
(7s6p6d1f) / [5s5p4d1f] Gaussian basis set for Au or the (7s12p6d1f) / [5s8p4d1f] one for At, respectively.
In the case of oxygen and hydrogen, triple-zeta contracted Gaussian basis sets were used~\cite{light_bas}.
The resulting basis was shown to be flexible enough for neglecting the basis set superposition errors.                                                        

\begin{figure}[htb]
\begin{minipage}[h!]{0.4\linewidth}
\center{\includegraphics[width=1\linewidth]{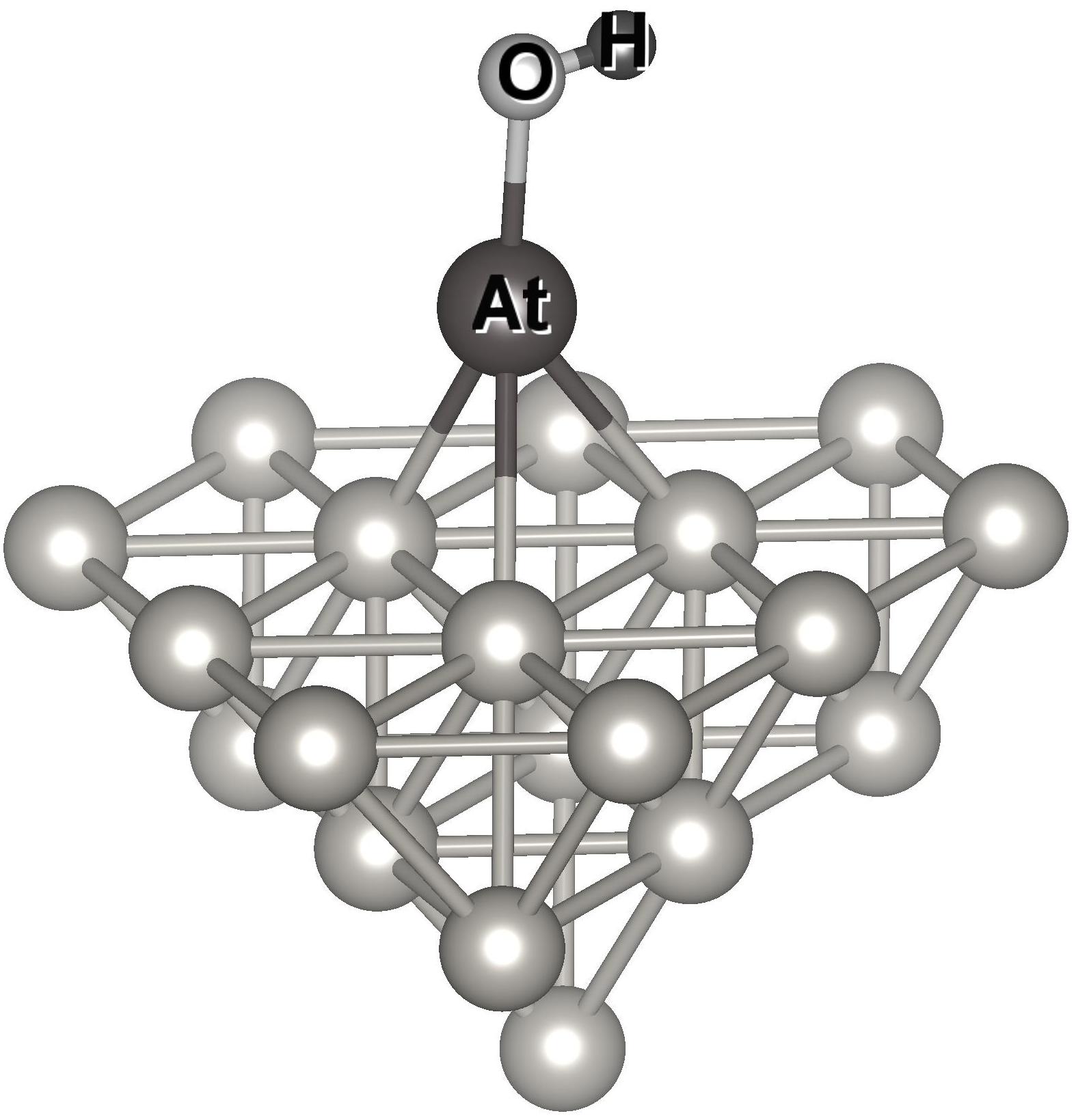}} \\
\end{minipage}
\hfill
\begin{minipage}[h!]{0.4\linewidth}
\center{\includegraphics[width=1\linewidth]{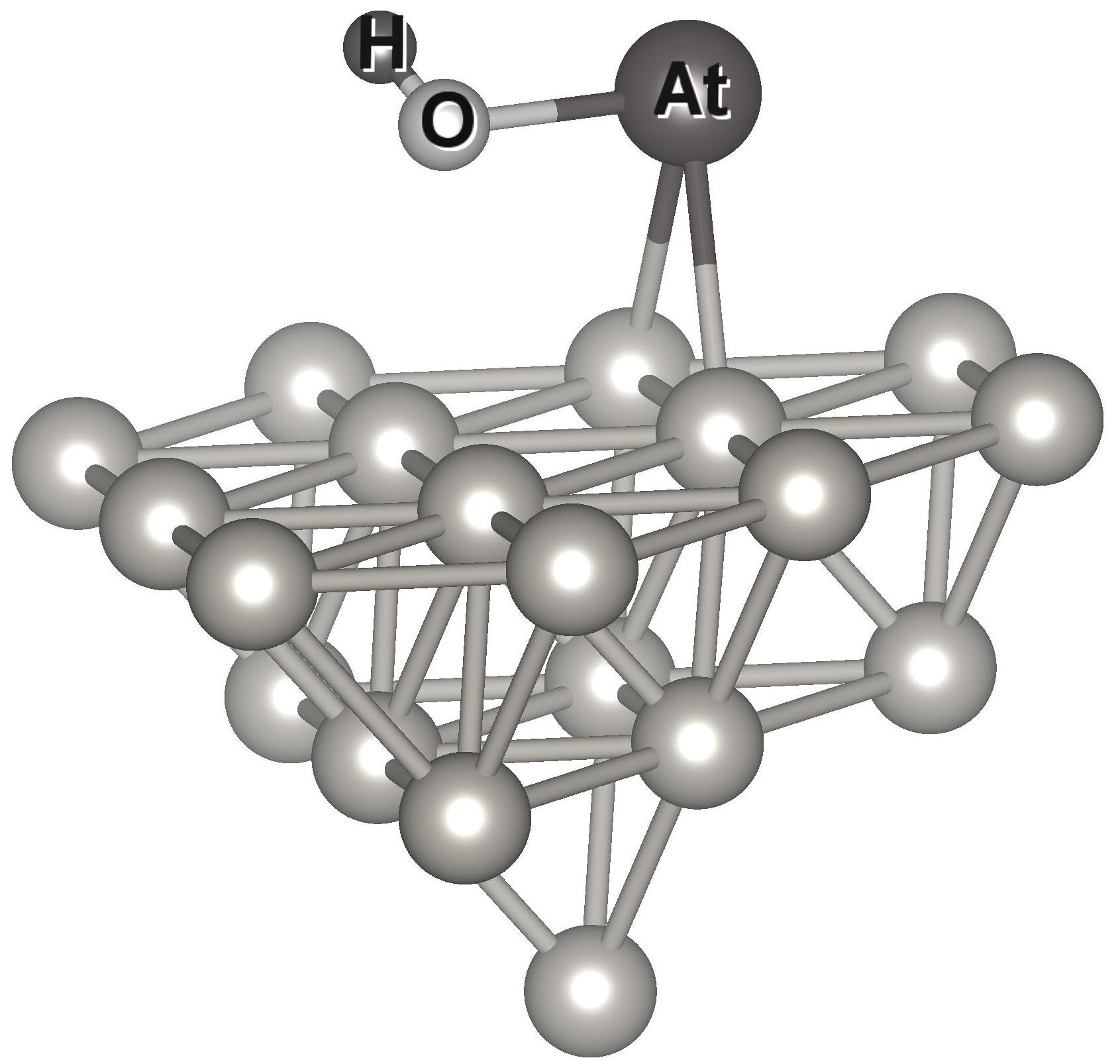}} \\
\end{minipage}
\vfill
\begin{minipage}[h!]{0.4\linewidth}
\center{lateral position}
\end{minipage}
\hfill
\begin{minipage}[h!]{0.4\linewidth}
\center{flat position}
\end{minipage}
\caption{
$\rm AtOH-Au_n$ complexes.
In the ``lateral'' case, the At atom is located in a ``hollow'' position, while
for the ``flat'' case, the oxygen atom is the nearest one to the cluster surface.    
\label{fgr:atoh_pos}
}
\end{figure}

Several gold clusters comprising up to 69 atoms were chosen to simulate the adsorption site of At/AtOH on the most stable gold (111) surface.
These clusters were chosen as fragments of an ideal gold crystal, with the Au--Au distances kept at their experimental crystal values. 
Figure~\ref{fgr:pos} illustrates the possible positions of the At atom on the surface of different gold clusters at the extremum points on the $\rm At-Au_n$ potential energy surface.
The respective equilibrium distance between the adsorbed At atom and the gold clusters was calculated separately for each of the shown adsorption complexes.
In the case of the adsorption of AtOH on a gold cluster the equilibrium positions of the molecule were determined by allowing the molecule to move over the cluster surface.
Based on a series of preliminary calculations, two competing positions for AtOH molecule on the Au cluster surface were found (see Fig.~\ref{fgr:atoh_pos}).
In the first case, the formation of a chemical bond between At and the gold atoms occurs, whereas in a second case the bonding is mainly associated with a charge redistribution on the cluster surface.
Since the stabilization of binding energies upon increasing the cluster size $n$
can be occasional, the convergence was additionally monitored by analyzing        
the Bader net charge~\cite{bader1,bader2} of the astatine atom as well as the equilibrium distance At--gold as functions of the cluster size.
The applicability and overall accuracy of the used basis sets and the calculation technique has been demonstrated in the works~\cite{120_our,e113oh_our}.

\begin{figure*}[h!]
\begin{minipage}[h!]{0.46\linewidth}
\center{on the top}
\end{minipage}
\hfill
\begin{minipage}[h!]{0.46\linewidth}
\center{bridge}
\end{minipage}
\vfill
\begin{minipage}[h!]{0.46\linewidth}
\center{\includegraphics[width=1\linewidth]{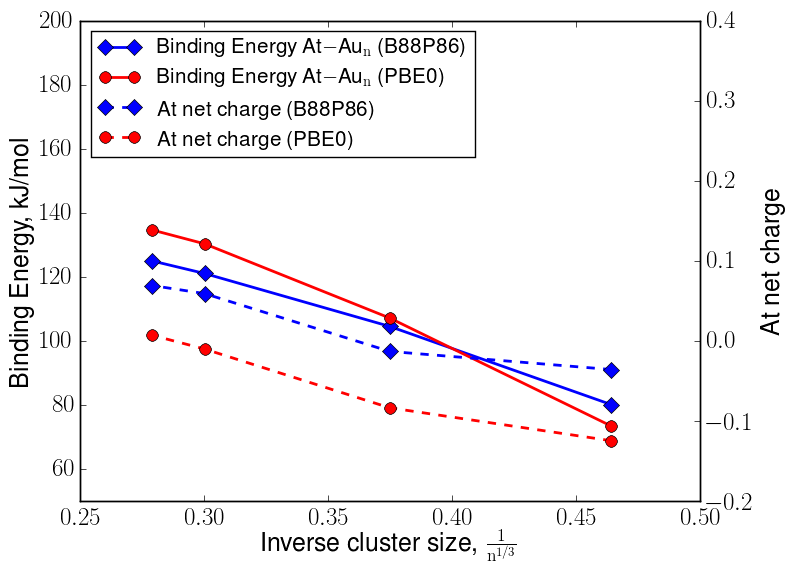}} \\
\end{minipage}
\hfill
\begin{minipage}[h!]{0.46\linewidth}
\center{\includegraphics[width=1\linewidth]{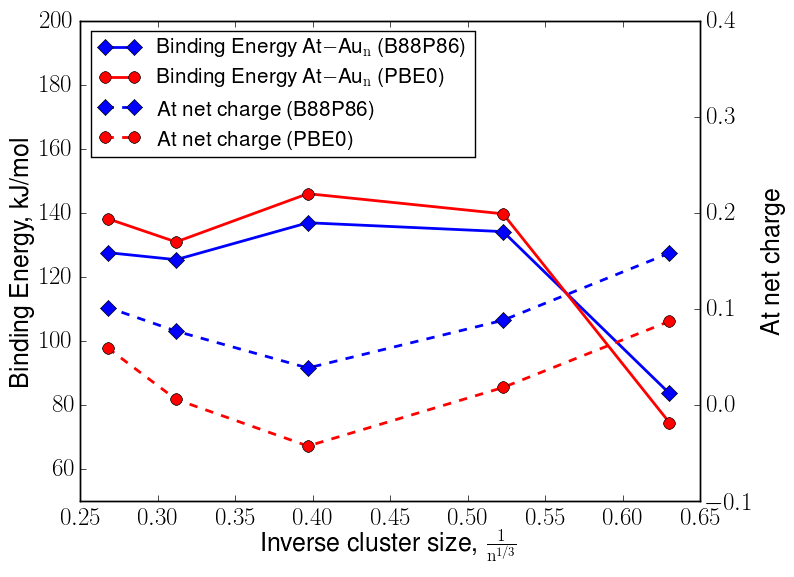}} \\
\end{minipage}
\vfill
\begin{minipage}[h!]{0.46\linewidth}
\center{hollow-2}
\end{minipage}
\hfill
\begin{minipage}[h!]{0.46\linewidth}
\center{hollow-3}
\end{minipage}
\vfill
\begin{minipage}[h!]{0.46\linewidth}
\center{\includegraphics[width=1\linewidth]{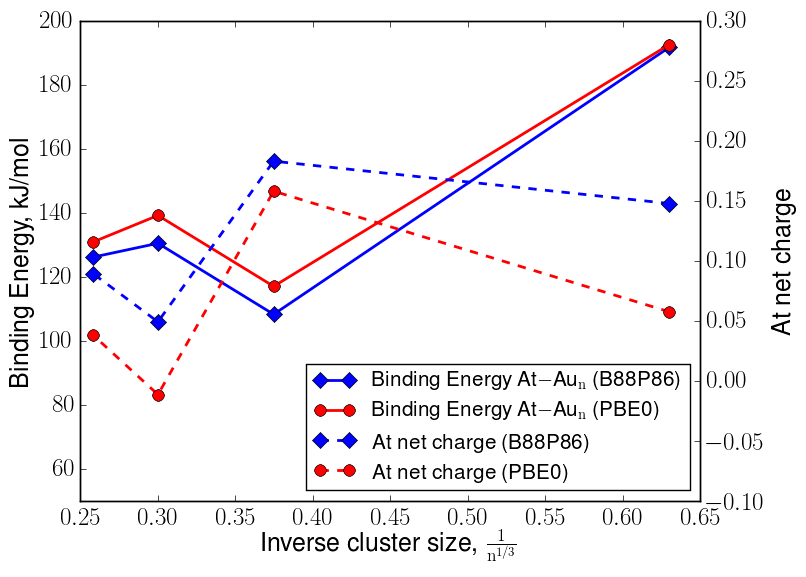}} \\
\end{minipage}
\hfill
\begin{minipage}[h!]{0.46\linewidth}
\center{\includegraphics[width=1\linewidth]{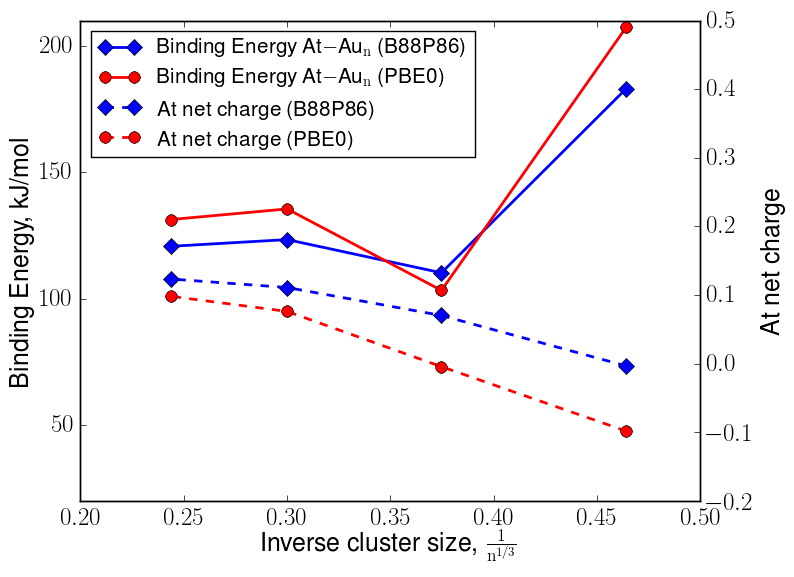}} \\
\end{minipage}
\caption{
A summary of the binding energies and the Bader net charges of At for the different $\rm At-Au_n$ adsorption complexes.
\label{fgr:ataun}
}
\end{figure*}
\section*{Results}
\subsection* {Structure of $\rm At-Au_n$ complexes}
The calculated bond length $R_e$ of AtAu molecule in the B88P86 approximation for the exchange-correlational functional is $R_e=2.644$~\AA,
while the corresponding value for the PBE0-approximation equals to $R_e=2.621$~\AA.
The obtained dissociation energy of the AtAu molecule is 190~kJ/mol for the B88P86 case and 180~kJ/mol for the PBE0 case. 
The chemical bond is formed without significant charge transfer.

For the more complex case of an adsorption of At on the surface of an Au cluster, Figure~\ref{fgr:ataun} illustrates the dependence of the calculated At--Au$\rm_n$ binding energies and the Bader net charges of At in the different generic positions as functions of the inverse cluster size. 
It can be seen that from a cluster size equal or larger than Au$\rm_{37}$ (i.e., less than 0.30 in terms of inverse cluster size) the binding energy of the At atom with gold cluster varies only slightly around 125~kJ/mol for B88P86 and 134~kJ/mol for PBE0. 
In all these complexes, the At atom bears a small positive Bader net charge. Furthermore, the astatine atom adsorbed in the bridge position on the surface of a gold cluster was found to be slightly more stable.
The equilibrium distances between the At atom and the surface layer of gold atoms are 2.8~\AA\, for the top position, 2.6~\AA\, for the bridge and hollow-2 positions, and 
2.5~\AA\, for an adsorption in hollow-3 site.

\begin{figure}[h]                                                               
\centering                                                                      
\includegraphics[height=6cm]{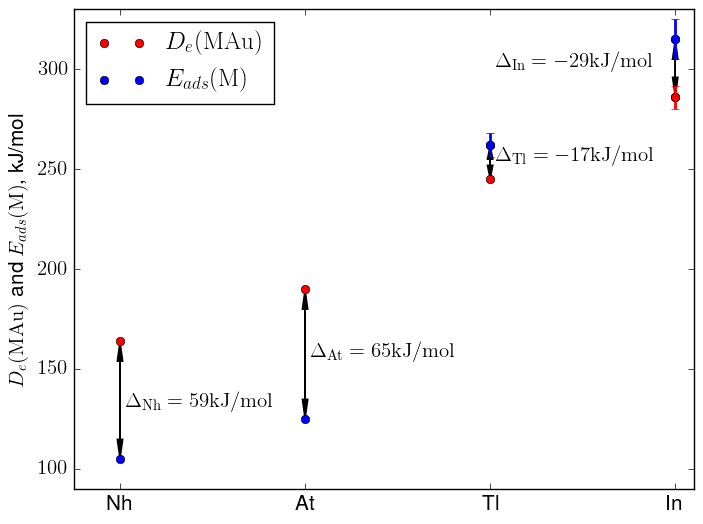}                                          
\caption{
Comparison of the dissociation energies $D_e$(MAu) [kJ/mol] of MAu molecules with the adsorption energies $E_{ads}$(M) [kJ/mol] of single M atoms on a gold surface, 
where M = Nh, At, Tl, and In. The corresponding differences in binding energies $\rm \Delta_M$ [kJ/mol] are indicated for each considered element. 
Experimental data has been retrieved from Refs.~\cite{at_gold, tl_aun, in_aun, handbook_new} and are presented with error bars, whereas the theoretical results were obtained within the 2c-RDFT (B88P86) approach for At (present work) and from~\cite{113_gold_our_old, rus113} for Nh and Tl.
\label{fgr:semiempirical}}
\end{figure}

The obtained information on the interaction of astatine with gold is sufficient for evaluating the adsorption energy for single atoms of elemental Nh on the same surface 
using semi-empirical relations~\cite{113oh,113_pershina,persh_rev}. These relations are based on the assumption, that for homologs M,
the difference between the dissociation energy of the MAu molecule and the binding energy of the corresponding $\rm M-Au_n$ system remains nearly the same, independent of the cluster size $n$ and the adsorption position.  
The available theoretical and experimental data for the interaction of Nh, its nearest homologs Tl and In, as well as its pseudo-homologue At with a gold surface has been compiled in Figure~\ref{fgr:semiempirical}.
The mentioned differences between the dissociation energy of a MAu molecule (with M = Nh and At) and the respective adsorption energy of single atoms of M on the surface of a gold cluster are stable and amount to approximately~60~kJ/mol.
For Tl and In, this assumption of ``constant differences'' holds too, however with differences of another magnitude and an opposite trend (i.e., -20~kJ/mol).
Based on these differences from the immediate homologs Tl and In, one obtains an adsorption energy for Nh atoms on gold of $E^{Au}_{ads}$(Nh) $\rm \approx$ 180~kJ/mol. This result coincides with previous semi-empirical evaluations (164~kJ/mol from~\cite{eichler:15} and 159~kJ/mol from~\cite{113oh}). 
However, the question arises, whether these results hold for adsorptions at high temperatures. 
According to available experimental data, the adsorption temperatures for Tl and In on a gold surface are 858$^{\circ}$C and 980$^{\circ}$C, respectively~\cite{tl_aun, in_aun}. 
Meanwhile, the corresponding temperature for At lies at a somewhat lower level of only 387$^{\circ}$C~\cite{at_gold}, which make these predictions possibly more trustworthy.
Nevertheless, the adsorption behavior of adatoms on a hot surface can be rather different from the adsorption on a cold one (see e.g. Ref.~\cite{nadine_new}).
As cluster simulations are generally based on the assumption, that the gold surface around the adsorption site is absolutely stable (i.e., $\rm T=0~K$), the direct applicability of gathered results for estimating adsorption energies at high temperatures is questionable.
Spontaneous breaking of atomic bonds on the thermally activated surface can even lead to the incorporation of adatoms into the surface layer.
Thus, for an accurate prediction of the adsorption behavior, particularly for Tl and In on a gold surface, further research is needed.

\begin{figure}[h]                                                               
\centering                                                                      
\includegraphics[height=6cm]{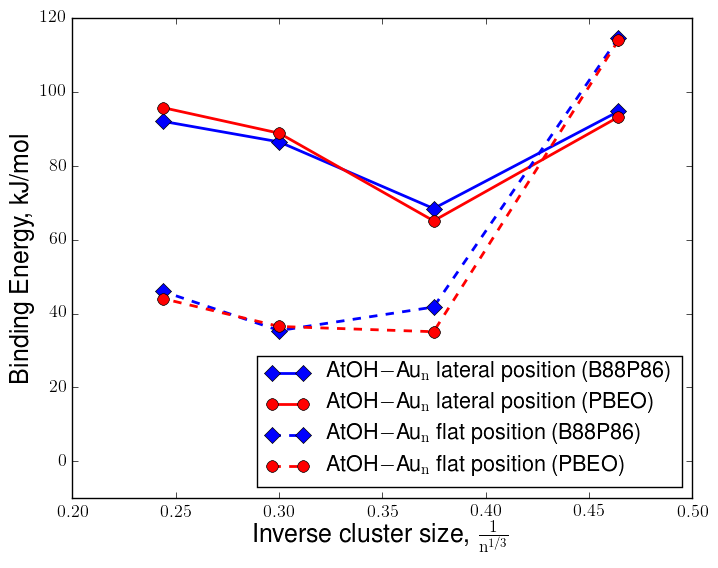}                                          
\caption{Binding energies of the astatine atoms for the lateral and flat positions in the $\rm AtOH-Au_n$ complexes. 
\label{fgr:atoh_aun}}
\end{figure}

\subsection* {Structure of $\rm AtOH-Au_n$ complexes}
The binding energy of an AtOH molecule with gold clusters consisting of more than 19 atoms (i.e., less than 0.37 in terms of inverse cluster size) in the flat position 
lies within the 50 -- 30~kJ/mol range (see also Fig.~\ref{fgr:atoh_pos}). This is substantially lower, than that obtained for the lateral AtOH position with binding energies
between 80 -- 100~kJ/mol (see Fig.~\ref{fgr:atoh_aun}).

In the flat position, the oxygen atom is the closest one to the gold surface. The associated equilibrium distance between oxygen and the nearest gold atom of the surface is 2.96~\AA.
Meanwhile, the At atom is located in the bridge position with equilibrium distances to the nearest gold atoms of 3.08~\AA\, and 3.18~\AA. In this case, chemical bonding without charge transfer takes place.
In the lateral position however, the At atom is situated above the gold clusters close to a hollow-3 position. Here, the determined equilibrium distances between the astatine atom and nearest gold atoms lie in the range  3.1 -- 3.2~\AA.
The At-O bond lengths in the $\rm AtOH-Au_n$ adsorption complexes is slightly larger than for AtOH case.
Generally, the charge transfer in $\rm AtOH-Au_n$ adsorption complexes is stronger than the one occurring in $\rm At-Au_n$; the gold clusters act as an electron donor.

\section*{Conclusions}
To summarize, extensive calculations on cluster models of At/AtOH--Au(111) adsorption complexes,
performed with accurate small-core relativistic pseudopotentials                
and RDFT correlation treatment, predict an adsorption energy of about
130$\rm \pm$10~kJ/mol for At--Au(111) and 90$\rm \pm$10~kJ/mol for AtOH--Au(111).
These values agree with the data based on experimental findings (i.e., $\rm 147\pm 15$~kJ/mol for At, and $\rm 100^{+20}_{-10}$~kJ/mol for AtOH~\cite{at_gold}), 
thus confirming the correctness of the interpretation of experimental data in the cited work.
The performed calculations confirm the experimental observation on the formation of AtOH molecules in presence of trace amounts of water and oxygen in the carrier gas.
Due to the previously discussed similarities~\cite{e113oh_our} in the chemical properties of AtOH and NhOH molecules, one may expect,
that the formation of NhOH is indeed possible under experimental conditions similar to those in the experiments on At.

\section*{Acknowledgements}
We would like to thank Prof. C. van W{\"u}llen for providing the relativistic DFT code. 
Thanks are due to Patrick Steinegger for helpful discussions.
This work was carried out using computing resources of the federal collective usage center «Complex for simulation and data processing for mega-science facilities» at NRC "Kurchatov Institute", http://ckp.nrcki.ru/
The work was partially supported by the by the Russian Science Foundation grant No. 14-31-00022.


\begin{thebibliography}{31}
\expandafter\ifx\csname natexlab\endcsname\relax\def\natexlab#1{#1}\fi
\expandafter\ifx\csname bibnamefont\endcsname\relax
  \def\bibnamefont#1{#1}\fi
\expandafter\ifx\csname bibfnamefont\endcsname\relax
  \def\bibfnamefont#1{#1}\fi
\expandafter\ifx\csname url\endcsname\relax
  \def\url#1{\texttt{#1}}\fi
\expandafter\ifx\csname urlprefix\endcsname\relax\def\urlprefix{URL }\fi
\providecommand*{\bibinfo}[2]{#2}
\providecommand*{\eprint}[1]{#1}
\providecommand*{\url}[1]{#1}
\begingroup\makeatletter
 \@temptokena{%
  \expandafter\ifx\csname citenamefont\endcsname\relax
   \DeclareRobustCommand\citenamefont{\@firstofone}%
   \global\let\citenamefont\citenamefont
   \global\expandafter\let\csname citenamefont \expandafter\endcsname\csname
  citenamefont \endcsname
  \fi
 }\if@filesw\immediate\write\@auxout{\the\@temptokena}\fi
\expandafter\endgroup\the\@temptokena

\bibitem[{\citenamefont{Oganessian and Utyonkov}(2015)}]{ogan_rew}
\bibinfo{author}{\bibfnamefont{Y.~T.} \bibnamefont{Oganessian}}
  \bibnamefont{and} \bibinfo{author}{\bibfnamefont{V.}~\bibnamefont{Utyonkov}},
  \bibinfo{journal}{Nucl. Phys. A} \textbf{\bibinfo{volume}{944}},
  \bibinfo{pages}{62} (\bibinfo{year}{2015}).

\bibitem[{\citenamefont{Herzberg} \emph{et~al.}(2006)\citenamefont{Herzberg,
  Greenlees, Butler, Jones, Venhart, Darby, Eeckhaudt, Eskola, Grahn,
  Gray-Jones} \emph{et~al.}}]{island}
\bibinfo{author}{\bibfnamefont{R.-D.} \bibnamefont{Herzberg}},
  \bibinfo{author}{\bibfnamefont{P.}~\bibnamefont{Greenlees}},
  \bibinfo{author}{\bibfnamefont{P.}~\bibnamefont{Butler}},
  \bibinfo{author}{\bibfnamefont{G.}~\bibnamefont{Jones}},
  \bibinfo{author}{\bibfnamefont{M.}~\bibnamefont{Venhart}},
  \bibinfo{author}{\bibfnamefont{I.}~\bibnamefont{Darby}},
  \bibinfo{author}{\bibfnamefont{S.}~\bibnamefont{Eeckhaudt}},
  \bibinfo{author}{\bibfnamefont{K.}~\bibnamefont{Eskola}},
  \bibinfo{author}{\bibfnamefont{T.}~\bibnamefont{Grahn}},
  \bibinfo{author}{\bibfnamefont{C.}~\bibnamefont{Gray-Jones}}, \emph{et~al.},
  \bibinfo{journal}{Nature}
  \textbf{\bibinfo{volume}{442}}(\bibinfo{number}{7105}), \bibinfo{pages}{896}
  (\bibinfo{year}{2006}).

\bibitem[{\citenamefont{Oganessian}(2017)}]{og_island}
in \bibinfo{author}{\bibfnamefont{Y.}~\bibnamefont{Oganessian}}\emph{\bibinfo{booktitle}{8-th
  Int. Particle Accelerator Conf.(IPAC'17), Copenhagen, Denmark, 14 -- 19 May,
  2017}} (\bibinfo{organization}{JACOW, Geneva, Switzerland},
  \bibinfo{year}{2017}), pp. \bibinfo{pages}{4848--4851}.

\bibitem[{\citenamefont{Karol} \emph{et~al.}(2016)\citenamefont{Karol, Barber,
  Sherrill, Vardaci, and Yamazaki}}]{IUPAC}
\bibinfo{author}{\bibfnamefont{P.~J.} \bibnamefont{Karol}},
  \bibinfo{author}{\bibfnamefont{R.~C.} \bibnamefont{Barber}},
  \bibinfo{author}{\bibfnamefont{B.~M.} \bibnamefont{Sherrill}},
  \bibinfo{author}{\bibfnamefont{E.}~\bibnamefont{Vardaci}}, \bibnamefont{and}
  \bibinfo{author}{\bibfnamefont{T.}~\bibnamefont{Yamazaki}},
  \bibinfo{journal}{Pure Appl. Chem.}
  \textbf{\bibinfo{volume}{88}}(\bibinfo{number}{1-2}), \bibinfo{pages}{139}
  (\bibinfo{year}{2016}).

\bibitem[{\citenamefont{Eichler} \emph{et~al.}(2007)\citenamefont{Eichler,
  Aksenov, Belozerov, Bozhikov, Chepigin, Dmitriev, Dressler, Gaeggeler, V.A.,
  Haenssler, Itkis, Laube} \emph{et~al.}}]{exp_112aun}
\bibinfo{author}{\bibfnamefont{R.}~\bibnamefont{Eichler}},
  \bibinfo{author}{\bibfnamefont{N.}~\bibnamefont{Aksenov}},
  \bibinfo{author}{\bibfnamefont{A.}~\bibnamefont{Belozerov}},
  \bibinfo{author}{\bibfnamefont{G.}~\bibnamefont{Bozhikov}},
  \bibinfo{author}{\bibfnamefont{V.}~\bibnamefont{Chepigin}},
  \bibinfo{author}{\bibfnamefont{S.}~\bibnamefont{Dmitriev}},
  \bibinfo{author}{\bibfnamefont{R.}~\bibnamefont{Dressler}},
  \bibinfo{author}{\bibfnamefont{H.}~\bibnamefont{Gaeggeler}},
  \bibinfo{author}{\bibfnamefont{G.}~\bibnamefont{V.A.}},
  \bibinfo{author}{\bibfnamefont{F.}~\bibnamefont{Haenssler}},
  \bibinfo{author}{\bibfnamefont{M.}~\bibnamefont{Itkis}},
  \bibinfo{author}{\bibfnamefont{A.}~\bibnamefont{Laube}},
  \bibinfo{author}{\bibfnamefont{V.}~\bibnamefont{Lebedev}},
  \bibinfo{author}{\bibfnamefont{O.}~\bibnamefont{Malyshev}},
  \bibinfo{author}{\bibfnamefont{Y.}~\bibnamefont{Oganessian}},
  \bibinfo{author}{\bibfnamefont{O.}~\bibnamefont{Petrushkin}},
  \bibinfo{author}{\bibfnamefont{D.}~\bibnamefont{Piguet}},
  \bibinfo{author}{\bibfnamefont{P.}~\bibnamefont{Rasmussen}},
  \bibinfo{author}{\bibfnamefont{S.}~\bibnamefont{Shishkin}},
  \bibinfo{author}{\bibfnamefont{A.}~\bibnamefont{Shutov}},
  \bibinfo{author}{\bibfnamefont{A.}~\bibnamefont{Svirikhin}},
  \bibinfo{author}{\bibfnamefont{E.}~\bibnamefont{Tereshatov}},
  \bibinfo{author}{\bibfnamefont{G.}~\bibnamefont{Vostokin}},
  \bibinfo{author}{\bibfnamefont{M.}~\bibnamefont{Wegrzecki}},
  \bibnamefont{and} \bibinfo{author}{\bibfnamefont{A.}~\bibnamefont{Yeremin}},
  \bibinfo{journal}{Nature} \textbf{\bibinfo{volume}{447}}, \bibinfo{pages}{72}
  (\bibinfo{year}{2007}).

\bibitem[{\citenamefont{Eichler} \emph{et~al.}(2010)\citenamefont{Eichler,
  Aksenov, Albin, Belozerov, Bozhikov, Chepigin, Dmitriev, Dressler, Gaeggeler,
  Gorshkov, and Henderson}}]{exp_114aun}
\bibinfo{author}{\bibfnamefont{R.}~\bibnamefont{Eichler}},
  \bibinfo{author}{\bibfnamefont{N.~V.} \bibnamefont{Aksenov}},
  \bibinfo{author}{\bibfnamefont{Y.~V.} \bibnamefont{Albin}},
  \bibinfo{author}{\bibfnamefont{A.~V.} \bibnamefont{Belozerov}},
  \bibinfo{author}{\bibfnamefont{G.~A.} \bibnamefont{Bozhikov}},
  \bibinfo{author}{\bibfnamefont{V.~I.} \bibnamefont{Chepigin}},
  \bibinfo{author}{\bibfnamefont{S.~N.} \bibnamefont{Dmitriev}},
  \bibinfo{author}{\bibfnamefont{R.}~\bibnamefont{Dressler}},
  \bibinfo{author}{\bibfnamefont{H.~W.} \bibnamefont{Gaeggeler}},
  \bibinfo{author}{\bibfnamefont{V.~A.} \bibnamefont{Gorshkov}},
  \bibnamefont{and}
  \bibinfo{author}{\bibfnamefont{G.}~\bibnamefont{Henderson}},
  \bibinfo{journal}{Radiochim. Acta}
  \textbf{\bibinfo{volume}{98}}(\bibinfo{number}{3}), \bibinfo{pages}{133}
  (\bibinfo{year}{2010}).

\bibitem[{\citenamefont{Dmitriev} \emph{et~al.}(2014)\citenamefont{Dmitriev,
  Aksenov, Albin, Bozhikov, Chelnokov, Chepigin, Eichler, Isaev, Katrasev,
  Lebedev, Malyshev, Petrushkin} \emph{et~al.}}]{e113_exp}
\bibinfo{author}{\bibfnamefont{S.~N.} \bibnamefont{Dmitriev}},
  \bibinfo{author}{\bibfnamefont{N.~V.} \bibnamefont{Aksenov}},
  \bibinfo{author}{\bibfnamefont{Y.~V.} \bibnamefont{Albin}},
  \bibinfo{author}{\bibfnamefont{G.~A.} \bibnamefont{Bozhikov}},
  \bibinfo{author}{\bibfnamefont{M.~L.} \bibnamefont{Chelnokov}},
  \bibinfo{author}{\bibfnamefont{V.~I.} \bibnamefont{Chepigin}},
  \bibinfo{author}{\bibfnamefont{R.}~\bibnamefont{Eichler}},
  \bibinfo{author}{\bibfnamefont{A.~V.} \bibnamefont{Isaev}},
  \bibinfo{author}{\bibfnamefont{D.~E.} \bibnamefont{Katrasev}},
  \bibinfo{author}{\bibfnamefont{V.~Y.} \bibnamefont{Lebedev}},
  \bibinfo{author}{\bibfnamefont{O.~N.} \bibnamefont{Malyshev}},
  \bibinfo{author}{\bibfnamefont{O.~V.} \bibnamefont{Petrushkin}},
  \bibinfo{author}{\bibfnamefont{L.~S.} \bibnamefont{Porobanuk}},
  \bibinfo{author}{\bibfnamefont{M.~A.} \bibnamefont{Ryabinin}},
  \bibinfo{author}{\bibfnamefont{A.~V.} \bibnamefont{Sabelnikov}},
  \bibinfo{author}{\bibfnamefont{E.~A.} \bibnamefont{Sokol}},
  \bibinfo{author}{\bibfnamefont{A.~V.} \bibnamefont{Svirikhin}},
  \bibinfo{author}{\bibfnamefont{G.~Y.} \bibnamefont{Starodub}},
  \bibinfo{author}{\bibfnamefont{I.}~\bibnamefont{Usoltsev}},
  \bibinfo{author}{\bibfnamefont{G.~K.} \bibnamefont{Vostokin}},
  \bibnamefont{and} \bibinfo{author}{\bibfnamefont{A.~V.}
  \bibnamefont{Yeremin}}, \bibinfo{journal}{Mendeleev Commun.}
  \textbf{\bibinfo{volume}{24}}(\bibinfo{number}{5}), \bibinfo{pages}{253}
  (\bibinfo{year}{2014}).

\bibitem[{\citenamefont{Aksenov} \emph{et~al.}(2017)\citenamefont{Aksenov,
  Steinegger, Abdullin, Albin, Bozhikov, Chepigin, Eichler, Lebedev, Madumarov,
  Malyshev} \emph{et~al.}}]{e113_exp_new}
\bibinfo{author}{\bibfnamefont{N.~V.} \bibnamefont{Aksenov}},
  \bibinfo{author}{\bibfnamefont{P.}~\bibnamefont{Steinegger}},
  \bibinfo{author}{\bibfnamefont{F.~S.} \bibnamefont{Abdullin}},
  \bibinfo{author}{\bibfnamefont{Y.~V.} \bibnamefont{Albin}},
  \bibinfo{author}{\bibfnamefont{G.~A.} \bibnamefont{Bozhikov}},
  \bibinfo{author}{\bibfnamefont{V.~I.} \bibnamefont{Chepigin}},
  \bibinfo{author}{\bibfnamefont{R.}~\bibnamefont{Eichler}},
  \bibinfo{author}{\bibfnamefont{V.~Y.} \bibnamefont{Lebedev}},
  \bibinfo{author}{\bibfnamefont{A.~S.} \bibnamefont{Madumarov}},
  \bibinfo{author}{\bibfnamefont{O.~N.} \bibnamefont{Malyshev}}, \emph{et~al.},
  \bibinfo{journal}{Eur. Phys. J. A}
  \textbf{\bibinfo{volume}{53}}(\bibinfo{number}{7}), \bibinfo{pages}{158}
  (\bibinfo{year}{2017}).

\bibitem[{\citenamefont{Zv{\'a}ra}(2008)}]{zvara}
\bibinfo{author}{\bibfnamefont{I.}~\bibnamefont{Zv{\'a}ra}},
  \emph{\bibinfo{title}{Experimental Developments in Gas-Phase Radiochemistry}}
  (\bibinfo{publisher}{Springer}, \bibinfo{year}{2008}).

\bibitem[{\citenamefont{Demidov and Zaitsevskii}(2014)}]{our_rew}
\bibinfo{author}{\bibfnamefont{Y.~A.} \bibnamefont{Demidov}} \bibnamefont{and}
  \bibinfo{author}{\bibfnamefont{A.}~\bibnamefont{Zaitsevskii}},
  \bibinfo{journal}{Russ. Chem. Bull.}
  \textbf{\bibinfo{volume}{63}}(\bibinfo{number}{8}), \bibinfo{pages}{1647}
  (\bibinfo{year}{2014}).

\bibitem[{\citenamefont{Rusakov} \emph{et~al.}(2013)\citenamefont{Rusakov,
  Demidov, and Zaitsevskii}}]{rus113}
\bibinfo{author}{\bibfnamefont{A.~A.} \bibnamefont{Rusakov}},
  \bibinfo{author}{\bibfnamefont{Y.~A.} \bibnamefont{Demidov}},
  \bibnamefont{and}
  \bibinfo{author}{\bibfnamefont{A.}~\bibnamefont{Zaitsevskii}},
  \bibinfo{journal}{Cent. Eur. J. Phys.}
  \textbf{\bibinfo{volume}{11}}(\bibinfo{number}{11}), \bibinfo{pages}{1537}
  (\bibinfo{year}{2013}).

\bibitem[{\citenamefont{Serov} \emph{et~al.}(2013)\citenamefont{Serov, Eichler,
  Dressler, Piguet, T{\"u}rler, V{\"o}gele, Wittwer, and
  G{\"a}ggeler}}]{tl_aun}
\bibinfo{author}{\bibfnamefont{A.}~\bibnamefont{Serov}},
  \bibinfo{author}{\bibfnamefont{R.}~\bibnamefont{Eichler}},
  \bibinfo{author}{\bibfnamefont{R.}~\bibnamefont{Dressler}},
  \bibinfo{author}{\bibfnamefont{D.}~\bibnamefont{Piguet}},
  \bibinfo{author}{\bibfnamefont{A.}~\bibnamefont{T{\"u}rler}},
  \bibinfo{author}{\bibfnamefont{A.}~\bibnamefont{V{\"o}gele}},
  \bibinfo{author}{\bibfnamefont{D.}~\bibnamefont{Wittwer}}, \bibnamefont{and}
  \bibinfo{author}{\bibfnamefont{H.}~\bibnamefont{G{\"a}ggeler}},
  \bibinfo{journal}{Radiochim. Acta}
  \textbf{\bibinfo{volume}{101}}(\bibinfo{number}{7}), \bibinfo{pages}{421}
  (\bibinfo{year}{2013}).

\bibitem[{\citenamefont{Demidov and Zaitsevskii}(2015)}]{e113oh_our}
\bibinfo{author}{\bibfnamefont{Y.}~\bibnamefont{Demidov}} \bibnamefont{and}
  \bibinfo{author}{\bibfnamefont{A.}~\bibnamefont{Zaitsevskii}},
  \bibinfo{journal}{Chem. Phys. Lett.} \textbf{\bibinfo{volume}{638}},
  \bibinfo{pages}{21 } (\bibinfo{year}{2015}), ISSN \bibinfo{issn}{0009-2614},

\bibitem[{\citenamefont{Zaitsevskii}
  \emph{et~al.}(2009)\citenamefont{Zaitsevskii, van W{\"u}llen, and
  Titov}}]{subperiodic}
\bibinfo{author}{\bibfnamefont{A.~V.} \bibnamefont{Zaitsevskii}},
  \bibinfo{author}{\bibfnamefont{C.}~\bibnamefont{van W{\"u}llen}},
  \bibnamefont{and} \bibinfo{author}{\bibfnamefont{A.~V.} \bibnamefont{Titov}},
  \bibinfo{journal}{Russ. Chem. Rev. (Engl. Transl.)}
  \textbf{\bibinfo{volume}{78}}(\bibinfo{number}{12}), \bibinfo{pages}{1173}
  (\bibinfo{year}{2009}).

\bibitem[{\citenamefont{Serov}
  \emph{et~al.}(2011{\natexlab{a}})\citenamefont{Serov, Aksenov, Bozhikov,
  Eichler, Dressler, Lebedev, Petrushkin, Piguet, Shishkin, Tereshatov,
  T{\"u}rler, Voegele} \emph{et~al.}}]{at_gold}
\bibinfo{author}{\bibfnamefont{A.}~\bibnamefont{Serov}},
  \bibinfo{author}{\bibfnamefont{N.}~\bibnamefont{Aksenov}},
  \bibinfo{author}{\bibfnamefont{G.}~\bibnamefont{Bozhikov}},
  \bibinfo{author}{\bibfnamefont{R.}~\bibnamefont{Eichler}},
  \bibinfo{author}{\bibfnamefont{R.}~\bibnamefont{Dressler}},
  \bibinfo{author}{\bibfnamefont{V.}~\bibnamefont{Lebedev}},
  \bibinfo{author}{\bibfnamefont{O.}~\bibnamefont{Petrushkin}},
  \bibinfo{author}{\bibfnamefont{D.}~\bibnamefont{Piguet}},
  \bibinfo{author}{\bibfnamefont{S.}~\bibnamefont{Shishkin}},
  \bibinfo{author}{\bibfnamefont{E.}~\bibnamefont{Tereshatov}},
  \bibinfo{author}{\bibfnamefont{A.}~\bibnamefont{T{\"u}rler}},
  \bibinfo{author}{\bibfnamefont{A.}~\bibnamefont{Voegele}},
  \bibinfo{author}{\bibfnamefont{D.}~\bibnamefont{Wittwer}}, \bibnamefont{and}
  \bibinfo{author}{\bibfnamefont{H.}~\bibnamefont{G{\"a}ggeler}},
  \bibinfo{journal}{Radiochim. Acta}
  \textbf{\bibinfo{volume}{99}}(\bibinfo{number}{9}), \bibinfo{pages}{593}
  (\bibinfo{year}{2011}{\natexlab{a}}).

\bibitem[{\citenamefont{Becke}(1988)}]{becke}
\bibinfo{author}{\bibfnamefont{A.~D.} \bibnamefont{Becke}},
  \bibinfo{journal}{Phys. Rev. A} \textbf{\bibinfo{volume}{38}},
  \bibinfo{pages}{3098} (\bibinfo{year}{1988}).

\bibitem[{\citenamefont{Perdew}(1986)}]{perdew}
\bibinfo{author}{\bibfnamefont{J.~P.} \bibnamefont{Perdew}},
  \bibinfo{journal}{Phys. Rev. B} \textbf{\bibinfo{volume}{33}},
  \bibinfo{pages}{8822} (\bibinfo{year}{1986}).

\bibitem[{\citenamefont{Adamo and Barone}(1999)}]{pbe0}
\bibinfo{author}{\bibfnamefont{C.}~\bibnamefont{Adamo}} \bibnamefont{and}
  \bibinfo{author}{\bibfnamefont{V.}~\bibnamefont{Barone}},
  \bibinfo{journal}{J. Chem. Phys.}
  \textbf{\bibinfo{volume}{110}}(\bibinfo{number}{13}), \bibinfo{pages}{6158}
  (\bibinfo{year}{1999}).

\bibitem[{\citenamefont{Zaitsevskii}
  \emph{et~al.}(2006)\citenamefont{Zaitsevskii, Rykova, Mosyagin, and
  Titov}}]{gibrid}
\bibinfo{author}{\bibfnamefont{A.}~\bibnamefont{Zaitsevskii}},
  \bibinfo{author}{\bibfnamefont{E.}~\bibnamefont{Rykova}},
  \bibinfo{author}{\bibfnamefont{N.~S.} \bibnamefont{Mosyagin}},
  \bibnamefont{and} \bibinfo{author}{\bibfnamefont{A.~V.} \bibnamefont{Titov}},
  \bibinfo{journal}{Cent. Eur. J. Phys.}
  \textbf{\bibinfo{volume}{4}}(\bibinfo{number}{4}), \bibinfo{pages}{448}
  (\bibinfo{year}{2006}).

\bibitem[{\citenamefont{Sch{\"a}fer}
  \emph{et~al.}(1994)\citenamefont{Sch{\"a}fer, Huber, and
  Ahlrichs}}]{light_bas}
\bibinfo{author}{\bibfnamefont{A.}~\bibnamefont{Sch{\"a}fer}},
  \bibinfo{author}{\bibfnamefont{C.}~\bibnamefont{Huber}}, \bibnamefont{and}
  \bibinfo{author}{\bibfnamefont{R.}~\bibnamefont{Ahlrichs}},
  \bibinfo{journal}{The Journal of Chemical Physics}
  \textbf{\bibinfo{volume}{100}}(\bibinfo{number}{8}), \bibinfo{pages}{5829}
  (\bibinfo{year}{1994}).

\bibitem[{\citenamefont{Sanville} \emph{et~al.}(2007)\citenamefont{Sanville,
  Kenny, Smith, and Henkelman}}]{bader1}
\bibinfo{author}{\bibfnamefont{E.}~\bibnamefont{Sanville}},
  \bibinfo{author}{\bibfnamefont{S.~D.} \bibnamefont{Kenny}},
  \bibinfo{author}{\bibfnamefont{R.}~\bibnamefont{Smith}}, \bibnamefont{and}
  \bibinfo{author}{\bibfnamefont{G.}~\bibnamefont{Henkelman}},
  \bibinfo{journal}{J. Comput. Chem.}
  \textbf{\bibinfo{volume}{28}}(\bibinfo{number}{5}), \bibinfo{pages}{899}
  (\bibinfo{year}{2007}).

\bibitem[{\citenamefont{Tang} \emph{et~al.}(2009)\citenamefont{Tang, Sanville,
  and Henkelman}}]{bader2}
\bibinfo{author}{\bibfnamefont{W.}~\bibnamefont{Tang}},
  \bibinfo{author}{\bibfnamefont{E.}~\bibnamefont{Sanville}}, \bibnamefont{and}
  \bibinfo{author}{\bibfnamefont{G.}~\bibnamefont{Henkelman}},
  \bibinfo{journal}{J. Phys.: Condens. Matter}
  \textbf{\bibinfo{volume}{21}}(\bibinfo{number}{8}), \bibinfo{pages}{084204}
  (\bibinfo{year}{2009}).

\bibitem[{\citenamefont{Demidov} \emph{et~al.}(2014)\citenamefont{Demidov,
  Zaitsevskii, and Eichler}}]{120_our}
\bibinfo{author}{\bibfnamefont{Y.~A.} \bibnamefont{Demidov}},
  \bibinfo{author}{\bibfnamefont{A.}~\bibnamefont{Zaitsevskii}},
  \bibnamefont{and} \bibinfo{author}{\bibfnamefont{R.}~\bibnamefont{Eichler}},
  \bibinfo{journal}{Phys. Chem. Chem. Phys.}
  \textbf{\bibinfo{volume}{16}}(\bibinfo{number}{6}), \bibinfo{pages}{2268 }
  (\bibinfo{year}{2014}).

\bibitem[{\citenamefont{Pershina} \emph{et~al.}(2009)\citenamefont{Pershina,
  Anton, and Jacob}}]{113oh}
\bibinfo{author}{\bibfnamefont{V.}~\bibnamefont{Pershina}},
  \bibinfo{author}{\bibfnamefont{J.}~\bibnamefont{Anton}}, \bibnamefont{and}
  \bibinfo{author}{\bibfnamefont{T.}~\bibnamefont{Jacob}},
  \bibinfo{journal}{Chem. Phys. Lett.}
  \textbf{\bibinfo{volume}{480}}(\bibinfo{number}{4}), \bibinfo{pages}{157}
  (\bibinfo{year}{2009}).

\bibitem[{\citenamefont{Pershina} \emph{et~al.}(2010)\citenamefont{Pershina,
  Borschevsky, Anton, and Jacob}}]{113_pershina}
\bibinfo{author}{\bibfnamefont{V.}~\bibnamefont{Pershina}},
  \bibinfo{author}{\bibfnamefont{A.}~\bibnamefont{Borschevsky}},
  \bibinfo{author}{\bibfnamefont{J.}~\bibnamefont{Anton}}, \bibnamefont{and}
  \bibinfo{author}{\bibfnamefont{T.}~\bibnamefont{Jacob}}, \bibinfo{journal}{J.
  Chem. Phys.} \textbf{\bibinfo{volume}{133}}, \bibinfo{pages}{104304}
  (\bibinfo{year}{2010}).

\bibitem[{\citenamefont{Pershina}(2015)}]{persh_rev}
\bibinfo{author}{\bibfnamefont{V.}~\bibnamefont{Pershina}},
  \bibinfo{journal}{Nucl. Phys. A} \textbf{\bibinfo{volume}{944}},
  \bibinfo{pages}{578} (\bibinfo{year}{2015}).

\bibitem[{\citenamefont{Rossbach and Eichler}(1984)}]{eichler:15}
\bibinfo{author}{\bibfnamefont{H.}~\bibnamefont{Rossbach}} \bibnamefont{and}
  \bibinfo{author}{\bibfnamefont{B.}~\bibnamefont{Eichler}},
  \emph{\bibinfo{title}{Adsorption of Metals on Metal Surfaces and the
  Possibilities of its Application in Nuclear Chemistry}},
  \bibinfo{type}{Report} ZFK-527, \bibinfo{institution}{Akademie der
  Wissenschaft der DDR} (\bibinfo{year}{1984}).

\bibitem[{\citenamefont{Serov}
  \emph{et~al.}(2011{\natexlab{b}})\citenamefont{Serov, Eichler, Dressler,
  Piguet, T{\"u}rler, V{\"o}gele, Wittwer, and G{\"a}ggeler}}]{in_aun}
\bibinfo{author}{\bibfnamefont{A.}~\bibnamefont{Serov}},
  \bibinfo{author}{\bibfnamefont{R.}~\bibnamefont{Eichler}},
  \bibinfo{author}{\bibfnamefont{R.}~\bibnamefont{Dressler}},
  \bibinfo{author}{\bibfnamefont{D.}~\bibnamefont{Piguet}},
  \bibinfo{author}{\bibfnamefont{A.}~\bibnamefont{T{\"u}rler}},
  \bibinfo{author}{\bibfnamefont{A.}~\bibnamefont{V{\"o}gele}},
  \bibinfo{author}{\bibfnamefont{D.}~\bibnamefont{Wittwer}}, \bibnamefont{and}
  \bibinfo{author}{\bibfnamefont{H.}~\bibnamefont{G{\"a}ggeler}},
  \bibinfo{journal}{Radiochim. Acta}
  \textbf{\bibinfo{volume}{99}}(\bibinfo{number}{2}), \bibinfo{pages}{95}
  (\bibinfo{year}{2011}{\natexlab{b}}).

\bibitem[{\citenamefont{Chiera} \emph{et~al.}(2017)\citenamefont{Chiera,
  Aksenov, Albin, Bozhikov, Chepigin, Dmitriev, Dressler, Eichler, Lebedev,
  Madumarov} \emph{et~al.}}]{nadine_new}
\bibinfo{author}{\bibfnamefont{N.}~\bibnamefont{Chiera}},
  \bibinfo{author}{\bibfnamefont{N.}~\bibnamefont{Aksenov}},
  \bibinfo{author}{\bibfnamefont{Y.}~\bibnamefont{Albin}},
  \bibinfo{author}{\bibfnamefont{G.}~\bibnamefont{Bozhikov}},
  \bibinfo{author}{\bibfnamefont{V.}~\bibnamefont{Chepigin}},
  \bibinfo{author}{\bibfnamefont{S.}~\bibnamefont{Dmitriev}},
  \bibinfo{author}{\bibfnamefont{R.}~\bibnamefont{Dressler}},
  \bibinfo{author}{\bibfnamefont{R.}~\bibnamefont{Eichler}},
  \bibinfo{author}{\bibfnamefont{V.~Y.} \bibnamefont{Lebedev}},
  \bibinfo{author}{\bibfnamefont{A.}~\bibnamefont{Madumarov}}, \emph{et~al.},
  \bibinfo{journal}{J. Radioanal. Nucl. Chem.}
  \textbf{\bibinfo{volume}{311}}(\bibinfo{number}{1}), \bibinfo{pages}{99}
  (\bibinfo{year}{2017}).

\bibitem[{\citenamefont{Haynes}(2014)}]{handbook_new}
\bibinfo{author}{\bibfnamefont{W.~M.} \bibnamefont{Haynes}},
  \emph{\bibinfo{title}{CRC Handbook of Chemistry and Physics}}
  (\bibinfo{publisher}{CRC press}, \bibinfo{year}{2014}).

\bibitem[{\citenamefont{Zaitsevskii}
  \emph{et~al.}(2011)\citenamefont{Zaitsevskii, Titov, Rusakov, and van
  W{\"u}llen}}]{113_gold_our_old}
\bibinfo{author}{\bibfnamefont{A.}~\bibnamefont{Zaitsevskii}},
  \bibinfo{author}{\bibfnamefont{A.~V.} \bibnamefont{Titov}},
  \bibinfo{author}{\bibfnamefont{A.~A.} \bibnamefont{Rusakov}},
  \bibnamefont{and} \bibinfo{author}{\bibfnamefont{C.}~\bibnamefont{van
  W{\"u}llen}}, \bibinfo{journal}{Chem. Phys. Lett.}
  \textbf{\bibinfo{volume}{508}}(\bibinfo{number}{4}), \bibinfo{pages}{329}
  (\bibinfo{year}{2011}).

\end{thebibliography}

\end{document}